\documentclass[twocolumn,preprintnumbers,amsmath,amssymb]{revtex4}

\usepackage{graphicx}
\usepackage{dcolumn}
\usepackage{bm}

\begin{document}

\title{Fresh look at the Hagedorn mass spectrum as seen in the experiments}

\author{K. A. Bugaev, V. K. Petrov  and G. M. Zinovjev}
\affiliation{Bogolyubov Institute for Theoretical Physics,
Kiev, Ukraine
}

\date{\today}
\begin{abstract}

The  medium dependent finite width is introduced  into an exactly solvable model 
with the general  mass-volume spectrum of the QGP bags. 
The model allows us  to estimate the minimal 
value  of the QGP bags' width  from the lattice QCD data.
The large width of the QGP bags  not only explains 
the observed deficit in the number of  hadronic resonances  {comparing  to the Hagedorn mass spectrum,} but also clarifies the reason   why 
the heavy QGP bags   cannot be directly observed  as metastable  states in a hadronic phase.  

\vspace*{0.25cm} 

\end{abstract}

\maketitle


\section{Introduction and putting forward the problem}
The statistical bootstrap model (SBM) \cite{Hagedorn:65}  
was the first to suggest 
that the exponentially increasing mass spectrum of hadrons, the Hagedorn spectrum,
could lead to new thermodynamics above the Hagedorn temperature $T_H$. Shortly after,  it has been demonstrated that both the dual resonance model (DRM) \cite{DRM, Miranski:73} (which originated the  string-like picture of hadrons) and the MIT bag model (which supposes the nontrivial vacuum structure) resemble the other features of SBM besides the asymptotic form of mass spectrum  \cite{MITBagM}.
Moreover, it has been realized that the Hagedorn temperature 
might be interpreted as the temperature of phase transition to the partonic degrees of freedom  \cite{Parisi:75}.
Henceforth these results initiated the extensive study of hadron 
thermodynamics within the model of a gas of bags (GBM) \cite{Kapusta:81}. 
The analytical solution
of GBM with a non-zero proper volume of hadronic bags (with the hard core repulsion)
allowed one to become aware of possible mechanism of deconfining phase transition from 
hadronic matter to the quark gluon plasma (QGP) (set by an infinite bag containing 
free quarks and gluons)  \cite{Gorenstein:81}.
Amazingly,  up to now GBM remains one of the most 
efficient phenomenological instruments to successfully describe the bulk properties 
of hadron production in existing experimental data on relativistic heavy ion 
collisions and due to the simplicity of its foundations to easily incorporate newly 
discovered features of strongly interacting matter   \cite{HG}.
Apparently, the most recent 
attempts to update GBM bringing the contemporary knowledge of the  phase diagram
of  quantum  chromodynamics (QCD)
 \cite{Bugaev:05c,Bugaev:07,CGreiner:06}
are entirely based on the lattice approach to quantum chromodynamics (LQCD)
\cite{fodorkatz, karsch}. \\
\indent
However, despite the considerable success of these models and their remarkable features all of them face two conceptual difficulties. The first one can be formulated by asking a very simple question: 'Why are the QGP bags never directly observed  in the experiments?' The routine argument applied to both high energy heavy ion and hadron collisions is that there exists  a phase transition and, hence, the huge energy gap separating the QGP bags from the ordinary (light) hadrons prevents the QGP co-existence  with the hadrons at  densities below the phase transition. The same line of  arguments is also valid  if the strong cross-over  exists. But on the other hand
in the laboratory experiments we are dealing with the finite systems and it is known  from the exact analytical solutions of the  constrained statistical multifragmentation model (SMM) \cite{Bugaev:04a}  and GBM \cite{Bugaev:05c} that there is a non-negligible probability to find the small and not too heavy QGP bags in thermally equilibrated finite systems  even in the confined (hadronic) phase. 
Therefore,  for finite volume systems  created in  high energy nuclear or   particle  collisions such QGP bags  could appear like any other  metastable states in statistical mechanics,
{since in this case  the statistical suppression is just a few orders  of magnitude and not of the order of   the Avogadro number.}
Moreover, at the pre-equilibrated stage of high energy collision nothing  can actually prevent their 
appearance. Then, if  such QGP bags  can be created there must be a reason which 
prevents their  direct  experimental  detection. 
{As we will show here there is an inherent property of the strongly 
interacting matter equation of state (EoS) 
which prevents their appearance 
inside of the  hadronic phase even in finite systems.  The same  property is also responsible for  the instability 
of  large or heavy  strangelets. 
}\\
\indent
The second conceptual  problem is rooted in   a huge deficit of  the  number of observed  hadronic resonances 
\cite{Bron:04}  with masses above 2.5 GeV predicted by the SBM and used, so far,  by all other subsequent  models discussed above. 
Thus, there is a paradox  situation with the Hagedorn mass  spectrum: it was predicted for heavy hadrons
which nowadays  must be regarded as QGP bags, but  
it can be experimentally  established up to hadronic masses of  about 2.3 GeV  \cite{Bron:04},
{ whereas  the recent review of Particle Data Group 
contains  very few  heavier hadronic resonances 
comparing to the SBM expectations.
Moreover, the best  description of particle yields observed in a very wide range of  
collision  energies of heavy ions   is  achieved 
by the statistical model which incorporates s all hadronic resonances not heavier than 2.3 GeV \cite{HG}.
Thus, it looks like  heavier hadronic species, except for the long living ones, are simply absent in 
the experiments \cite{Blaschke:03}. 
}
Of course, one could  argue that heavy hadronic resonances cannot be established experimentally 
because both  their   large width  and  very large number of decay channels lead to  great  difficulties in their identification, but the point is that, except for  the recent efforts  \cite{Blaschke:03},  the influence of   large  width of heavy 
resonances  on their EoS  properties and  the corresponding  experimental consequences 
were  rather  not studied in full. \\
\indent
Therefore, here we would like to introduce the  finite and medium dependent width of QGP bags into the statistical 
model, 
 study  its impact  on the   pressure of system at  zero baryonic density and  show 
that  the {\it subthreshold suppression of the QGP bags}  of this finite width model (FWM)  resolves  both  the conceptual problems 
discussed
above.  Our  aim is to make a firm  bridge between the statistical description of QGP,
the LQCD results  and the  most general  properties of  
{ hadronic 
mass spectrum as seen in experiments at high energies. 
}


\section{Basic Ingredients of the  FWM}
The most convenient way to study the phase structure  of  any statistical  model similar to the SBM, GBM or  the QGP bags with surface tension model (QGBSTM) 
\cite{Bugaev:07}  implies to use the isobaric partition \cite{Gorenstein:81,Bugaev:07, Bugaev:04a} and find its rightmost singularities. Hence,  after the Laplace transform  the  FWM grand canonical  partition  $Z(V,T)$ generates the following 
isobaric partition:
\begin{eqnarray}\label{Zs}
\hspace*{-0.25cm}\hat{Z}(s,T) \equiv \int\limits_0^{\infty}dV\exp(-sV)~Z(V,T) =\frac{1}{ [ s - F(s, T) ] } \,,
\end{eqnarray}

\vspace*{-0.05cm}
\noindent
where the function $F(s, T)$ includes the discrete $F_H$ and continuous $F_Q$ mass-volume spectrum 
of the bags 
\vspace*{-0.3cm}
\begin{eqnarray}
F(s,T)&\equiv& F_H(s,T)+F_Q(s,T) = \sum_{j=1}^n g_j e^{-v_js} \phi(T,m_j) 
\nonumber 
\\
 %
&
+&\int\limits_{V_0}^{\infty}dv\hspace*{-0.1cm}\int\limits_{M_0}^{\infty}
 \hspace*{-0.1cm}dm~\rho(m,v)\exp(-sv)\phi(T,m)~.
 \label{FsHQ}
\end{eqnarray}

\vspace*{-0.25cm}
\noindent
The bag density
 of mass $m_k$, eigen volume $v_k$  and degeneracy factor $g_k$
is given by  $\phi_k(T) \equiv g_k ~ \phi(T,m_k) $  with 
%
$$
\phi_k(T)   \equiv  \frac{g_k}{2\pi^2} \int\limits_0^{\infty}\hspace*{-0.1cm}p^2dp~
e^{\textstyle - \frac{(p^2~+~m_k^2)^{ \frac{1}{2} }}{T} }
=  g_k \frac{m_k^2T}{2\pi^2}~{ K}_2 {\textstyle \left( \frac{m_k}{T} \right) }\, .
$$ 
 
\vspace*{-0.15cm}
\noindent
The mass-volume spectrum $\rho(m,v)$  generalizes   the exponential mass spectrum 
introduced by Hagedorn \cite{Hagedorn:65}.  As  in the GBM and QGBSTM,  the 
FWM bags 
are assumed to have the hard core repulsion of the Van der Waals type  generating  the suppression factor proportional to the  exponential of  bag 
proper  volume $\exp(-sv)$. 
The first term of Eq.~(\ref{FsHQ}), $F_H$, represents the contribution of a finite number of low-lying
hadron states up to mass $M_0 \approx 2 $ GeV \cite{Goren:82}. 
$F_H$ 
has no $s$-singularities at
any temperature $T$ and  generates  a simple pole  (\ref{Zs})
that describes a hadronic phase, whereas  
we will prove that 
the mass-volume spectrum of the bags $F_Q(s,T)$  
leads to  an essential  singularity $s_Q^* (T) \equiv p_Q(T)/T$ which defines  the QGP  pressure $p_Q(T)$  at zero baryonic densities 
\cite{Gorenstein:81,Goren:82, Bugaev:07}. 
Any   singularity  
$s^*$ of $\hat{Z}(s,T)$ (\ref{Zs}) 
is defined by the equation    $s^*(T)~=~ F(s^*,T)$ \cite{Gorenstein:81,Bugaev:07}.

Here we use the simplest parameterization of  the  spectrum $\rho(m,v)$ to demonstrate the idea.
{ Additional  physical  justification of  the FWM
along with the analysis of  the FWM relation to the Regge trajectories of heavy QGP bags and  the corresponding  experimental consequences can be found in Refs. \cite{Reggeons:08} and \cite{Bugaev:08}, respectively.
We, however,  stress that}
the requirements discussed in the introduction do not leave
us too much freedom to construct such a spectrum.
Thus, to have a firm bridge with the most general experimental and theoretical findings of 
particle phenomenology 
it is necessary to 
assume that the continuous  hadronic mass spectrum has a Hagedorn like form
%
\begin{eqnarray}\label{Rfwm}
& \rho (m,v) =   \frac{ \rho_1 (v)  ~N_{\Gamma}}{\Gamma (v) ~m^{a+\frac{3}{2} } }
 \exp{ \textstyle \left[ \frac{m}{T_H}   -   \frac{(m- B v)^2}{2 \Gamma^2 (v)}  \right]  } \,, \\ 
&  \rho_1 (v) = f (T)\, v^{-b}~ \exp{\textstyle \left[  -  \frac{\sigma(T)}{T} \, v^{\kappa}\right] }\,.
\label{R1fwm}
 \end{eqnarray}
Also  this spectrum has 
 the Gaussian attenuation  around the bag mass
$B v$ 
 determined by
 the volume dependent  Gaussian  width $\Gamma (v)$ or width hereafter. 
We will distinguish it from the true width defined as 
$\Gamma_R = \alpha \, \Gamma (v)$ ($\alpha \equiv 2 \sqrt{2 \ln 2}\,$).

In practice  for narrow resonances  there used  two mass distributions, the Breit-Wigner 
and the Gaussian ones. As  will be shown below 
the Gaussian dependence is of a crucial importance  for the FWM because 
the Breit-Wigner attenuation leads to a divergency of the partition function. 
This is quite different from the early attempts to consider the  width of QGP bags in 
\cite{Blaschke:03}.

The normalization factor in (\ref{Rfwm}) is defined to 
obey the condition
\begin{eqnarray}\label{Ng}
& N_{\Gamma}^{-1}~ = ~ \int\limits_{M_0}^{\infty}
 \hspace*{-0.1cm} \frac{dm}{\Gamma(v)}
    \exp{\textstyle \left[  -   \frac{(m- B v)^2}{2 \Gamma^2 (v)}  \right] } \,.
 \end{eqnarray}
 %
  
It is important that the volume spectrum in  (\ref{R1fwm}) contains the surface free energy (${\kappa} = 2/3$) with the $T$-dependent 
surface tension which is parameterized by 
$\sigma(T) = \sigma_0 \cdot
\left[ \frac{ T_{c}   - T }{T_{c}} \right]^{2k + 1} $  ($k =0, 1, 2,...$) \cite{Bugaev:07, Bugaev:04b},
where  $ \sigma_0 > 0 $ could, nevertheless,   be a smooth function of temperature. 
In \cite{Bugaev:07} it is shown  that  such a  parameterization of  the bag surface tension 
is of  crucial importance  to generate the QCD tricritical endpoint.
For $T$ not above  the tricritical temperature $T_{c}$  this  form of $\sigma(T)$  is justified by the usual  cluster models 
like the Fisher droplet model  \cite{Fisher:67} and SMM \cite{Bondorf:95, Bugaev:00}, whereas 
the general   $T$ dependence    can be analytically derived from the surface partitions of the Hills and Dales model 
\cite{Bugaev:04b}. 
The important   consequences of such a
surface tension   and a discussion   of the curvature free energy 
absence in 
(\ref{R1fwm}) can be found in~\cite{Bugaev:07, Complement}.

An attempt of Ref. \cite{Goren:82} to derive  the bag pressure \cite{MITBagM}  within  the 
GBM  {is based on  a complicated  mathematical construct, but does not 
 explain an underlying  physical reason for the mass-volume 
spectrum of bags suggested in  \cite{Goren:82}.
In contrast to \cite{Goren:82}, 
the spectrum (\ref{Rfwm}) (and (\ref{R1fwm})) 
is  simple, but  general and adequate   for the medium dependence  of
both the width $\Gamma (v)$ and the bag's mass density $B$.}
It clearly reflects the fact 
that the QGP bags are similar to   the ordinary  quasiparticles with the medium dependent characteristics (life-time, most probable values of  mass and volume). 
Now we are ready to  
derive the  pressure  of an infinite bag
for two dependencies: the volume independent width $\Gamma(v) = \Gamma_0$ and 
the volume dependent width as $\Gamma(v) = \Gamma_1 \equiv \gamma v^\frac{1}{2}$.

\section{Analysis of the  FWM spectrum}
First we note that for large bag volumes ($v \gg M_0/B > 0$) the factor (\ref{Ng})  can be
found as  $N_\Gamma \approx 1/\sqrt{2 \pi} $.  Similarly, one can show that  for heavy free bags  ($m \gg B V_0$, $V_0 \approx 1$ fm$^3$ \cite{Goren:82},
{ignoring the  hard core repulsion and thermostat})
%
\begin{eqnarray}\label{Rm}
& \rho(m)  ~ \equiv   \int\limits_{V_0}^{\infty}\hspace*{-0.1cm} dv\,\rho(m,v) ~\approx ~
\frac{  \rho_1 (\frac{m}{B}) }{B ~m^{a+\frac{3}{2} } }
\exp{ \textstyle \left[ \frac{m}{T_H}     \right]  } \,,
\end{eqnarray}

\vspace*{-0.05cm}
\noindent
i.e. the  spectrum (\ref{Rfwm})
integrated over the bag  volume has a Hagedorn form modified by the surface free energy. 
It results from the fact that  for heavy bags the 
Gaussian  in (\ref{Rfwm}) acts as  a Dirac $\delta$-function for
either choice of $\Gamma_0$ or $\Gamma_1$. 
Thus, the Hagedorn form of  (\ref{Rm}) receives  a clear physical meaning and gives an additional argument in favor of the FWM. Also it gives an upper bound for the 
volume dependence of $\Gamma(v)$: the Hagedorn-like mass spectrum (\ref{Rm}) can be derived, if for large $v$ the width  $\Gamma$ increases  slower than $v^{(1 - \kappa/2)}= v^{2/3}$. 

Similarly to Eq. (\ref{Rm}), one can estimate the width of heavy free bags  averaged over their  volumes and get  $ \overline{\Gamma(v) } \approx  \Gamma(m/B) $.
Thus, 
with choosing   $\Gamma(v) = \Gamma_1(v)$ the mass spectrum of heavy free QGP bags 
is found to be  the Hagedorn-like one and  heavy resonances 
 develop 
the large  mean width $ \Gamma_1(m/B) = \gamma \sqrt{m/B}$. Hence,  they 
are hard to be observed. 
{Applying these arguments to the strangelets,
we conclude  that, if their mean volume is a few cubic fermis or larger, they  should survive for a  very short time,
which is in line with the results of \cite{Strangelets:06}.

Note also that such a mean width is essentially different from both the linear mass dependence of string models  \cite{StringW} and from an  exponential  form  of the nonlocal field theoretical models \cite{NLFTM}.}

Next we  calculate  $F_Q(s,T)$ (\ref{FsHQ}) for the  spectrum (\ref{Rfwm}) performing the mass integration. There are two distinct 
options 
 depending on the sign of the most probable mass: 
\begin{eqnarray}\label{Mprob}
& \langle m \rangle ~ \equiv ~  B v + \Gamma^2 (v) \beta\,,\quad {\rm with} 
\quad \beta \equiv  T_H^{-1} - T^{-1} \,. 
\end{eqnarray}
If {\boldmath 
$ \langle m \rangle > 0$} for $v \gg V_0$,  one can use the saddle point method
for mass integration to  find  the function~$F_Q (s,T)$
\begin{eqnarray}\label{FQposM}
&  F_Q^+ (s,T)   \approx \left[  \frac{T}{2\pi} \right]^{\frac{3}{2} }
\int\limits_{V_0}^{\infty}dv ~ \frac{ \rho_1(v) }{\langle m \rangle^a} ~\exp{\textstyle \left[  \frac{(p^+  - sT )v}{T}  \right]} \, 
\end{eqnarray}

\vspace*{-0.05cm}
\noindent
and the pressure of large  bags 
$p^+ \equiv T \left[ \beta B + \frac{\Gamma^2 (v)}{2 v} \beta^2 \right]$.
To get  (\ref{FQposM}) one has to employ  in (\ref{FsHQ}) an asymptotics  of the $K_2$-function  $\phi(T,m)\simeq (mT/2\pi)^{3/2}\exp(-m/T)$ for $m\gg~T$, 
{collect all $m$-dependent terms 
in exponential, get a full square for $(m -  \langle m \rangle)$ 
} and 
perform 
the Gaussian integration.

Since for $s  <  s_Q^*(T) \equiv p^+(v\rightarrow \infty)/T $ the integral  (\ref{FQposM}) diverges on its upper limit, 
the  partition (\ref{Zs}) has  an essential singularity corresponding to 
the QGP pressure {inside}  of  an  infinite   bag. 
It allows one  to conclude 
the width  $\Gamma$ cannot increase  faster than $v^{1/2}$ for $v\rightarrow \infty$, otherwise $p^+(v\rightarrow \infty) \rightarrow \infty $ and  $F_Q^+ (s,T)$ diverges for any $s$.
Thus,  for {\boldmath $ \langle m \rangle > 0$} the phase structure of the FWM  with  $\Gamma (v) \neq 0$ is similar to the QGBSTM \cite{Bugaev:07}.

The bag spectrum  $F_Q^+ (s,T)$ (\ref{FQposM}) is of general nature  and, 
{unlike} the suggestion  of \cite{Goren:82},  has a transparent  physical origin. One can also see that  two general  sources  of the bulk part of  bag free energy 
\begin{eqnarray}\label{PposM}
&  
- p^+ v = - T \left[ \beta\, \langle m \rangle - \frac{1}{2}\,\Gamma^2 (v) \beta^2 \right]
\end{eqnarray}

\vspace*{-0.05cm}
\noindent
are the bag most  probable   mass  and its width. 
Different $T$ dependent functions $\langle m \rangle$ and 
$\Gamma^2 (v)$ 
lead to 
different 
EoS. 

If  instead of  the Gaussian width parameterization  in
(\ref{Rfwm}) we used  the Breit-Wigner one, then we would not be able to derive 
the continuous spectrum $F_Q^+ (s,T)$ (\ref{FQposM})  and the corresponding bag pressure for any  nonvanishing bag  width $\Gamma (v)$. Indeed,  for $T > T_H$ the mass integrals in  $F_Q (s,T)$
would diverge like in SBM, unless the Breit-Wigner mass attenuation has a zero width or an exponentially increasing width 
$\Gamma \sim \exp [m /T_H ]$   \cite{Blaschke:03}.
The former does not resolve the both of the GBM conceptual problems,
whereas the latter corresponds to a very specific  ansatz  for the resonance  width which is in contradiction with the FWM assumptions. 

It is  possible to use the spectrum (\ref{FQposM}) not only for infinite system volume but for 
finite volumes $V \gg V_0$ as well. In this case the upper limit of integration should be replaced by finite $V$ 
(see Ref. \cite{Bugaev:05c} for details).  It changes  the singularities of partition 
 (\ref{Zs}) to a set of simple poles  $ s_n^*(T)$ in the complex $s$-plane which are  defined by the same equation as for 
 $V \rightarrow \infty$.  Similarly to the finite $V$ solution of the GBM 
\cite{Bugaev:05c},  it can be shown that for finite $T$ the FWM  simple poles may have  a small positive or even negative real part which would lead to a non-negligible contribution of the QGP bags into the  spectrum  $F(s,T)$  (\ref{FsHQ}).
Thus,
if the spectrum (\ref{FQposM})  was the only volume spectrum of the QGP bags, then there would exist the non-negligible probability of finding   heavy QGP bags ($m \gg M_0$)  in finite systems  at 
 $T \ll T_H$.  
Therefore, using the results of   the finite volume GBM and SMM,  we  conclude that the spectrum 
(\ref{FQposM}) itself 
cannot  explain  the absence of  the QGP bags at  $T \ll T_H$  and, hence, an alternative explanation of this fact is required. 

Such an explanation corresponds to {the negative  values of}   {\boldmath$ \langle m \rangle \le 0 $} for $v \gg V_0$.
From (\ref{Mprob}) one can see that  
for the volume dependent width $\Gamma (v) = \Gamma_1 (v) $ the most probable mass $ \langle m \rangle $ inevitably becomes negative at low $T$, if $0 < B < \infty$. 
{Using the asymptotics of the $K_2$-function for large and small values of $\frac{m}{T}$ one can show that 
at low $T$  the maximum of the  Gaussian mass distribution is located  at 
$ \langle m \rangle \le 0$. 
Hence only the tail  of  
the  Gaussian mass distribution  close to $M_0$  contributes to $F_Q(s,T)$. 
By the steepest descent method and with the $K_2$-asymptotic form   for $M_0 T^{-1} \gg 1$ one gets}
\begin{eqnarray}\label{FQnegM}
\hspace*{-0.5cm}F_Q^-(s,T) \hspace*{-0.05cm} \approx  \hspace*{-0.05cm} \left[  \frac{T}{2\pi} \right]^{\hspace*{-0.05cm}\frac{3}{2} } 
\hspace*{-0.1cm}
\int\limits_{V_0}^{\infty} \hspace*{-0.05cm} dv  \frac{ \rho_1(v) N_{\Gamma}\, \Gamma (v)\, \exp{\textstyle \left[  \frac{(p^-  - sT )v}{T}  \right]}
}{M_0^a\,  \,[M_0 - \langle m \rangle  + a \, \Gamma^{2} (v)/ M_0 ]} \hspace*{-0.3cm}
\end{eqnarray}

\vspace*{-0.05cm}
\noindent
with the analytic form for the QGP bag pressure  
\begin{eqnarray}\label{pnegM}
p^-\big|_{v \gg V_0}  = {\textstyle  \frac{T}{v} \left[  \beta M_0 -  \frac{(M_0 - Bv)^2}{2\, \Gamma^{2} (v)}  
 \right] }\,.
\end{eqnarray}

\vspace*{-0.05cm}
\noindent
We would like to stress   the last result requires $B>~0$ and  cannot be generated  by  a weaker $v$-dependence  than  $\Gamma(v) = \Gamma_1(v)$. 
Indeed, if $B<0$, then the normalization factor (\ref{Ng}) would not be $1/\sqrt{2 \pi}$, but changes to 
$N_{\Gamma} \approx   [M_0 - \langle m \rangle]\, \Gamma^{-1} (v) \exp{\textstyle 
\left[    \frac{(M_0 - Bv)^2}{2\, \Gamma^{2} (v)}  \right]} $ and, thus,  it would cancel 
the leading  term in  pressure (\ref{pnegM}). Note that  the  inequality 
{\boldmath$ \langle m \rangle \le 0 $} for all $v \gg V_0$ with  $B > 0$ and 
finite $p^-(v \rightarrow \infty)$ is valid 
for  $\Gamma(v) = \Gamma_1(v)$ only.
{ The negative value of  $ \langle m \rangle $ is  an indicator of a different  physical
situation comparing to $ \langle m \rangle > 0$, but    has no physical meaning since 
$ \langle m \rangle \le 0 $ does not enter   the main physical observable  $p^- $.
}

The new outcome of this case with $B>0$ is that for $T < T_H$ the spectrum 
(\ref{FQnegM}) contains the lightest QGP bags having the  smallest volume since 
every term in the pressure (\ref{pnegM}) is negative.  The finite volume of the system is no longer  important   because only  the  smallest bags survive in (\ref{FQnegM}).
Moreover, if such bags are created, they would have the  masses  of about  $M_0$ and
the widths of  about $\Gamma_1(V_0)$, and, hence, they would hardly  be distinguishable 
from the usual low-mass hadrons. 
Thus, the situation   {\boldmath$ \langle m \rangle \le 0 $} with 
$B>0$ leads to the {\it subthreshold suppression of the QGP bags} at low temperatures,
since their most  likely  mass is below the mass threshold  $M_0$ of the spectrum $F_Q(s,T)$.  {Note that such an effect cannot be derived within  any of  the GBM-kind models  proposed earlier.}\\
\indent
The results  {received}   give us a unique opportunity to make a bridge between the
particle phenomenology, some experimental facts and LQCD 
{conclusions. For instance, if the most probable mass of the QGP bags is known along with the QGP pressure, one can estimate the width of these  bags directly from Eqs. (\ref{PposM})
and  (\ref{pnegM}). 
The FWM pressure depends on two functions, therefore, in order 
to find them it is necessary to  know the form of the QGP pressure  somewhere  in the  hadronic phase.  Unfortunately,  the present  LQCD data do not provide 
us with such a detailed information and, hence, at the moment  some additional 
assumptions are inevitable.  
To demonstrate the new possibilities of FWM
}
let us consider 
several examples of the  
QGP  EoS  and relate them  to the above results. \\
\indent
First, we study the possibility to get  the MIT bag model pressure 
 $p_{bag} \equiv \sigma T^4 - B_{bag} $  \cite{MITBagM}  by  the  stable QGP bags, i.e. 
$\Gamma (v) \equiv 0$. Equating the pressures $p^+$ and $p_{bag}$, one finds that 
$T_H$
must be related to a bag constant  as
$B_{bag} \equiv \sigma T^4_H$.  Then the mass density of such bags 
$\frac{\langle m \rangle}{v} \equiv  B  =  \sigma T_H (T  + T_H)(T^2 + T_H^2)$ is
always positive. Thus, the MIT bag model  EoS  can be easily obtained by 
the FWM approach, but, as  discussed earlier, such bags should have been  observed. \\ \indent
Secondly,  we consider the stable bags, $\Gamma (v) \equiv 0$, but without the Hagedorn 
spectrum, i.e. $T_H \rightarrow \infty$. 
Matching $p^+ = - B $ and $p_{bag}$,  we find that at low temperatures 
the bag mass density  $\frac{\langle m \rangle}{v} = B$ is positive, whereas for 
high $T$ the mass density cannot be positive and, hence,  
one cannot reproduce $p_{bag}$ as $B \le 0$ and  the resulting pressure is not $p^-$ (\ref{pnegM}),
but a zero, as seen from (\ref{FQnegM}), (\ref{pnegM}) and   $N_{\Gamma}$ expression for the limit $\Gamma (v) \rightarrow 0$. 
One can try to reproduce $p_{bag}$ with the finite $T$ dependent  width $\Gamma (v) = 2 \, \sigma T^5 v$ for  $T_H \rightarrow \infty$. Then one can get  $p_{bag}$ from 
$p^+$, but only for low temperatures obeying the inequality 
$\frac{\langle m \rangle}{v} = B_{bag} - 2\, \sigma T^4 > 0$. Thus,  these  two 
examples  teach  us that without the Hagedorn mass spectrum one cannot  get 
the MIT bag model pressure. \\
\indent
It is also possible to reproduce an alternative QGP  EoS 
$p_a = \sigma T^4 - A_1 T + A_0 $ ($ A_1 > 0$, $ A_0 \ge 0$) \cite{GorMog} even 
for  $\Gamma (v) \equiv 0$. 
{The linear  $T$-dependence in the  QGP pressure, which clearly has nonperturbative nature,  is seen \cite{Reggeons:08} both in  old  \cite{LQCD:1,LQCD:2} and 
fresh \cite{LQCD:3} LQCD data.
}
Choosing $T_H$ to be a positive solution of equation 
$A_0 = A_1 T_H - \sigma T_H^4 $, one obtains $p_a$ from $p^+$ for the 
mass density of bag 
$\frac{\langle m \rangle}{v} \equiv  B  =  \sigma T T_H (2\,T^2 + T T_H + T_H^2) - A_0$. 
If $A_0 = 0$ (found in \cite{GorMog, Reggeons:08} from the LQCD data \cite{LQCD:1,LQCD:2,LQCD:3}),  FWM is able to reproduce
$p_a$ for any  $T$, whereas for $A_0 > 0$ it 
works for temperatures 
obeying $B > 0$.  \\
\indent
Also the model with  the linear  $T$-dependent pressure $p_a$
and   $A_0 = 0$  allows us to  estimate roughly 
the width  $\Gamma_1 (V_0)$ in a  FWM. 
{
Matching $p_a$  with $p^- (v \rightarrow \infty) = - T \frac{B^2}{2\, \gamma^2}$
one can  determine  $B/\gamma$ ratio for $T \le c_\pm\, T_H$
($0 < c_\pm < 1$). 
For $T = 0$ one finds $A_1 = \frac{B_0^2}{2\, \gamma_0^2}$, i.e.  the FWM  naturally explains the linear $T$-dependent term in QGP pressure for  the nonvanishing bag  width coefficient $\gamma_0$ at $T=0$.\\
\indent
Putting  $p^+$ and $p^-(v \rightarrow \infty)$ equal   and solving for $\gamma^2$, one can find the switching value of the width coefficient $\gamma^2_\pm = - \frac{B}{ \beta}$ for the switching temperature $T_\pm = c_\pm T_H$. 
From this  result one can show that $0 < c_\pm < 1$ 
due to   the inequalities $B > 0$ and $\gamma^2_\pm >0$.\\
\indent
Then matching  $p_a$ and $p^+$, one obtains the width coefficient  
\begin{eqnarray}\label{gamma2}
&  
\gamma^2 = 2 \, \beta^{-1} [ \sigma T_H T (T^2 + TT_H + T_H^2) - B(T)] 
\end{eqnarray}
for $T \ge c_\pm\, T_H$.
Obviously, if  $(T-T_H)$ is 
an exact  divisor  of the difference  in (\ref{gamma2}), 
then 
$\gamma^2 > 0$  for all temperatures in the range $ c_\pm T_H \le  T \le T_H$.
}
The simplest possibility to obey such a requirement  is to assume  that $B(T) =  \sigma T_H^2  (T^2 + TT_H + T_H^2)$ for any $T$.
Then one gets 
 $ \gamma_0^2 = B_0^2 / (2 A_1) = T_H B_0 / 2 = \sigma T_H^5 /2$ for $T= 0$  and  
 $\gamma^2 = 2 \, T B(T) $ for $T \ge T_H$. 
Taking  the constants  in $T_c$  
units ($T_c \approx 200$ MeV), we obtain 
the true width for the SU(3) color group with two flavors 
\cite{LQCD:2} 
as
$\Gamma_R (V_0, T=0) \approx 1.22\,  V_0^{\frac{1}{2}} \, T_c^{\frac{5}{2}} \alpha \approx 587 $ MeV and
$\Gamma_R (V_0, T=T_H)   = \sqrt{12}\,\Gamma_R (V_0, T=0) \approx 2034$ MeV. 
This estimate clearly demonstrates us  that there is no way to detect the decays of such shortly living  QGP bags in the laboratory. 
A detailed analysis  of these findings  and  their sensitivity to 
different LQCD data  is presented in  \cite{Reggeons:08}.

\section{Conclusions}
Here we develop   the  novel statistical approach to study  the QGP bags with   medium dependent finite width. 
The  FWM   is based on the Hagedorn-like mass spectrum of bags modified by the surface free energy of bags and by the bag  width. We show  that the volume dependent width of the QGP bags $\Gamma (v) = \gamma\, v^\frac{1}{2}$ leads to the Hagedorn mass spectrum of free heavy bags.  Such a behavior of a width allows us to explain a  deficit of heavy hadronic resonances in the experimentally observed  mass spectrum. 
Under the  plausible  assumptions we derive  the general form   for  the bag pressure $p^+$ which accounts for the effect of finite  width in the EoS. We argue  that the obtained spectrum 
itself 
cannot  explain the absence of directly observable QGP bags in the high energy nuclear 
and  particle collisions. Then we find out a new possibility to ``hide''  the heavy  QGP bags for $T \ll T_H$  by their {\it subthreshold suppression}. The latter occurs due to the fact 
that at low  $T$ the most probable mass of heavy bags $\langle m \rangle \le 0$  
and, thus, is below the lower cut-off  $M_0$ of
the continuous mass spectrum. 
Hence only the lightest  bags of mass about $M_0$ and of  smallest volume $V_0$
may contribute into the resulting spectrum, but such QGP bags will be indistinguishable  
from the low-lying  hadronic resonances. 
We 
demonstrate 
how the FWM can reproduce a few EoS on the QGP market and corroborate that the low $T$ pressure $p^-$ reproduces properly some nonperturbative features revealed by LQCD. 
{ 
In principle, the FWM allows one to extract  the QGP pressure from the LQCD pressure for hadronic phase, 
if the contribution of  the discrete part of  hadronic mass-volume spectrum is also known.
The generalization to  non-zero baryonic densities is straightforward by assuming 
the dependence of the model functions $B$ and $\Gamma_1 (v)$  on the baryonic 
chemical potential. 
Our estimate of the volume dependent width looks 
 pretty encouraging for heavy ion phenomenology.
A detailed discussion of the FWM  experimental consequences   will be presented   elsewhere \cite{Reggeons:08,
Bugaev:08}. 
To simultaneously determine the most probable mass of the QGP bags and their width,   it would be nice  to study the metastable 
branch of the QGP EoS at low $T$  with the LQCD and compare its results  with the FDM. 
}

{\bf Acknowledgments.} We are thankful  to P. Braun-Munzinger, P. Giubellino, H. Gutbrot,  S. Molodtsov, D. H.  Rischke,
J. Stachel, H. St\"ocker and D. Voskresensky  for fruitful discussions and
 for 
important comments.
One of us, K.A.B., thanks the department KP1 of GSI, Darmstadt,  for a warm hospitality.



\begin{thebibliography}{99}

\bibitem{Hagedorn:65}
%
R. Hagedorn, Suppl. Nuovo Cimento {\bf 3}, 147  (1965).  

\bibitem{DRM}
%
K.~Huang and S.~Weinberg,
Phys.\ Rev.\ Lett.\  {\bf 25}, 895 (1970).

\bibitem{Miranski:73}
%
V.~A.~Miranskii, V.~P.~Shelest, B.~V.~Struminskii and G.~M.~Zinovjev,
Phys.\ Lett.\  B {\bf 43}, 73 (1973).

\bibitem{MITBagM}
%
A. Chodos {\it et. al.},  Phys. Rev. {\bf D 9},    3471  (1974). 

\bibitem{Parisi:75}
%
N. Cabibbo and G. Parisi, Phys. Lett. {\bf B 59},  67  (1975).


\bibitem{Kapusta:81}
%
J. I. Kapusta, Phys. Rev. {\bf D 23},   2444 (1981).


\bibitem{Gorenstein:81}
%
M. I. Gorenstein, V. K. Petrov and G. M. Zinovjev,
Phys. Lett. {\bf  B 106},  327 (1981).
%


\bibitem{HG}
%
for  more references see a review 
P. Braun-Munzinger, K. Redlich and J. Stachel, nucl-th/0304013, 109 p. 


\bibitem{Bugaev:05c}
%
K. A. Bugaev,  Phys.\ Part.\ Nucl. {\bf 38}, 447 (2007).


\bibitem{Bugaev:07}
K. A. Bugaev,
Phys. Rev. C {\bf 76},   014903 (2007) 
and  arXiv:hep-ph/0711.3169.

\bibitem{CGreiner:06}
%
for a review see  I. Zakout, C. Greiner, J. Schaffner-Bielich,
Nucl. Phys. {\bf A 781},  150 (2007).

\bibitem{fodorkatz}
%
Z. Fodor,
PoS {\bf LATTICE2007}, 011 (2007). 


\bibitem{karsch}
%
F. Karsch,
Prog.\ Theor.\ Phys.\ Suppl.\  {\bf 168}, 237 (2007).


\bibitem{Bugaev:04a}
%
K. A. Bugaev, 
 Acta. Phys. Polon. B {\bf 36},   3083 (2005).


\bibitem{Bron:04}
%
%
for a  discussion see W. Broniowski, W. Florkowski and L. Y. Glozman,
Phys. Rev. {\bf D 70}, 117503 (2004). 

\bibitem{Blaschke:03}
%
D. B. Blaschke and K. A. Bugaev,
 Fizika {\bf B 13},   491 (2004); 
%
Phys.\ Part.\ Nucl.\ Lett.\  {\bf 2},  305 (2005).
 

\bibitem{Goren:82}
%
M. I. Gorenstein, G. M. Zinovjev, V. K. Petrov, and V. P. Shelest, 
Teor. Mat. Fiz.  {\bf 52},  346 (1982).


\bibitem{Reggeons:08}
%
K. A. Bugaev, V. K. Petrov and G. M. Zinovjev,
{arXiv:0807.2391} [hep-ph] (2008) 13p.

\bibitem{Bugaev:08}
%
K. A. Bugaev, 
{arXiv:0809.1023} [nucl-th] (2008) 6p.


\bibitem{Bugaev:04b}
%
K. A. Bugaev, L. Phair and J. B. Elliott,
Phys. Rev. {\bf E 72},  047106  (2005);
%
K. A. Bugaev  and J. B. Elliott,
Ukr. J. Phys. {\bf 52} (2007) 301.

\bibitem{Fisher:67}
M. E. Fisher, Physics {\bf 3},  255 (1967).


\bibitem{Bondorf:95}
J. P. Bondorf {\it et al.},
Phys. Rep. {\bf 257},  131 (1995).

\bibitem{Bugaev:00}
K. A. Bugaev,
M. I. Gorenstein, I. N. Mishustin and W. Greiner,
Phys. Rev. {\bf C 62},  044320 (2000);
Phys. Lett. {\bf B 498},  144 (2001);
P. T. Reuter and K. A. Bugaev,
Phys. Lett. {\bf B 517} (2001) 233
and references therein.

\bibitem{Complement}
%
%
L. G. Moretto et al., 
 Phys. Rev. Lett. {\bf 94},  202701 (2005).

\bibitem{Strangelets:06} 
%
S. Yasui and A. Hosaka,
Phys. Rev. {\bf D 74}, 054036 (2006).
 
\bibitem{StringW} 
%
see, for instance, 
I. Senda, 
Z. Phys. {\bf C 55}, 331 (1992). 

\bibitem{NLFTM}
%
A. C. Kalloniatis, S. N. Nedelko and L. Smekal,
Phys. Rev. {\bf D 70}, 094037 (2004) and references therein.

\bibitem{GorMog}
%
M. I. Gorenstein and O. A. Mogilevsky, Z. Phys. {\bf C 38},163 (1988).

\bibitem{LQCD:1}
%
J. Engels, F. Karsch, J. Montway and H. Satz,
Nucl. Phys. {\bf B 205},  545 (1982).

\bibitem{LQCD:2}
%
T. Celik, J. Engels  and H. Satz,
Nucl. Phys. {\bf B 256},  670 (1985).

\bibitem{LQCD:3}
%
M. Cheng {et al.,} 
Phys. Rev. {\bf D 77},  014511 (2008).


\end{thebibliography}
\end{document}